# Arc plasma ablation of quartz crystals

Carles Corbella[1,*], Sabine Portal[1] and Michael Keidar[1]

[1] Department of Mechanical & Aerospace Engineering, George Washington University, 800 22nd Street, Northwest, Washington, DC 20052, United States of America

[*] E-mail: ccorberoc@gwu.edu

## Abstract

Spherical quartz stones of around 1 cm in diameter have been exposed to anodic arc discharges in a helium atmosphere at 300 Torr. The arc current flowing between the graphite electrodes was set either in continuous DC mode (30-150 A) or in pulsed mode at 2 Hz (220 A peak). The ablation rate in each sample was systematically measured after several seconds of arc plasma treatment. Optical emission spectroscopy (OES) diagnostics and 2-D fluid simulations of the arc discharge have shed light on the heat flux transport and the heating mechanisms of the quartz crystals. A linear correlation is found between the absorbed power density and the resulting rate of penetration, which yields a maximal value of 15 cm/h for approximately 150 W/cm$^2$. The linear fit on the slope provides a specific energy of 40 kJ/cm$^3$. The incident energy flux onto the sample surface promoted a phase transition from crystalline to glassy silica, as characterized via Raman spectroscopy. This study points out the strong potential of arc plasma technology for geothermal drilling applications.

# 1. Introduction

Arc technology encompasses a large number of applications, from material cutting and welding through synthesis of thin coatings and of nanomaterials [1-4]. Anodic arc discharge is an atmospheric arc plasma process in which an anode, made of a refractory material, is ablated due to large electron currents and high gas temperatures reached in the arc column, of several thousands Kelvin [5]. Besides applications in materials production and processing, such powerful energy source may also be adequate to modify surfaces of hard materials, like granite rocks. A similar project has been successfully deployed in the framework of laser-induced plasma ablation for precision-machining of quartz crystals and glass [6,7]. Zhang *et al* showed a correlation between the ablation rate and the laser fluence. In another example, plasma pretreatment was used to condition granite before cutting [8]. This conditioning step reduced dramatically the specific cutting energy of the rock. The specific energy of vaporization, S.E., is defined as the amount of energy irradiating the sample required to remove a unit volume of material, and it is a very useful concept to discuss the efficiency of geothermal drilling methods. Woskov and Cohn analyzed the directed energy millimeter wave (MMW) rock drilling technique, operated between 30 and 300 GHz, based on the measured penetration rate into granite, basalt, sandstone and limestone [9]. DC plasma torches have been also successfully adapted to treat natural stones and to process reinforced concrete and steel plates in different atmospheres, like argon and water vapour [10-12].

In this study, atmospheric anodic arc using graphite electrodes is proposed as an alternative energy resource candidate for efficient cutting and ablation of hard rocks. Quartz was selected as testing sample because of its ubiquity among the Earth minerals and its relatively high melting point (1950 K) and boiling point (2500 K). First, the arc plasma was characterized by OES, and the energy transport modes were studied via numerical simulations of the plasma parameters using a 2-D computational fluid dynamics code. Second, quartz crystals were placed in the vicinity of atmospheric arc discharges. The discharges were fed with DC currents of up to 150 A. In addition, pulsed arc within the Hz range, which is a novel nanosynthesis mode introduced elsewhere [13,14], has provided arc current peaks of approximately 200 A at duty cycles short enough to maintain power levels comparable to DC plasmas. A series of arc ablation tests were performed on stones of 10 mm or 14 mm in diameter. The penetration rate of the plasma energy input has been quantified as function of the power density, and it has been used to evaluate the potential of atmospheric arc discharge in drilling of mineral samples. The chemical structure of the treated samples was characterized by means of Raman spectroscopy to assess phase transitions and presence of impurities upon arc exposure. Finally, the characteristics of the arc ablation process are discussed and contrasted with geothermal drilling figures.

## 2. Methods

### 2.1. Experimental details

The characteristics of the experimental setup have been thoroughly described elsewhere [13,15]. In brief, the arc plasma ablation experiments were performed in a stainless-steel cylindrical vessel of 4500 $cm^3$ in volume, which contained two vertically, axially aligned electrodes made of graphite. The upper electrode (cathode, 10 mm in diameter) was fixed, while the lower electrode (anode, 3 mm in diameter) could be displaced up and down using a computer-controlled linear drive. The anodic arc discharge was ignited upon separation of the electrodes energized at a given arc current. Either DC power or pulsed power was provided by means of a Miller Gold Star 300SS DC power supply. DC arc current ranged from 30 to 150 A. Pulsed arc discharges were programmed via an external waveform generator, resulting in arc pulses at 2 Hz and 10% duty cycle (pulse duration: 50 ms) with 220 A of peak current and 50 V peak voltage. All processes were conducted at average powers of between 0.5 and 10 kW. Further details concerning the electrical setup for pulsed discharges are reported by Corbella *et al* [13]. The system was pumped down to a base pressure lower than 0.1 Torr using a rotary pump. A working pressure of 300 Torr in He (purity: 99.995%) was set by adjusting a throttle valve to the pump.

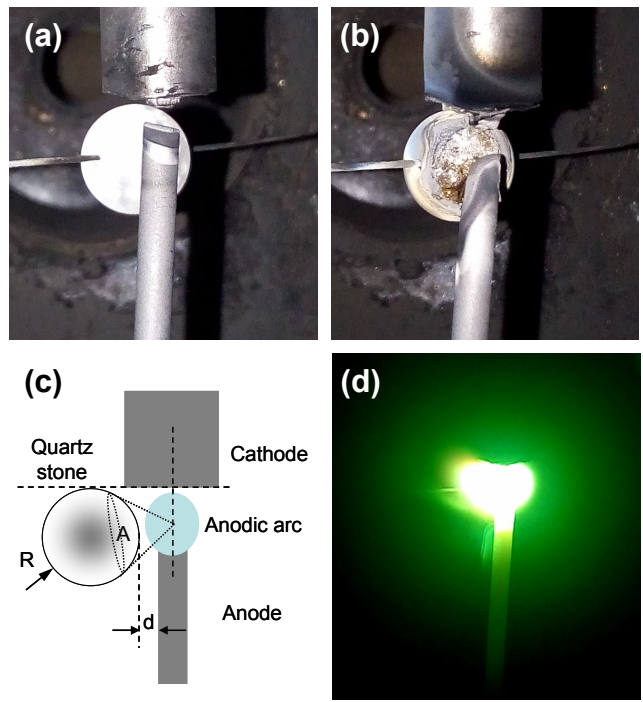

**FIG 1:** Experimental setup with a spherical quartz stone (radius $R$=5 mm) suspended, using a tungsten wire, at a distance $d$=2 mm to the anode surface. Images of the quartz stone **(a)** before and **(b)** after an arc discharge conducted at 300 Torr He with an arc current of 75 A (Sample 4). **(c)** Sketch of the experimental setup showing the relative positions of the graphite electrodes and the quartz stone. The plasma energy

emerging from the arc core is projected onto the area *A* of the stone. **(d)** Photograph (time resolution: 70 ms) of the arc process on Sample 4

Spherical stones consisting of quartz crystals (10 mm and 14 mm in diameter) were used to test the mineral evaporation efficiency of atmospheric anodic arcs. Such stones are labelled Sample 0 through 7. As shown in Fig. 1a, the stones were placed next to the anode by means of an horizontally oriented tungsten wire. The gap between the stone surface and the anode surface was $d$=2 mm. Shorter distances were not considered to avoid welding of the sample with the electrode. Tests at larger distances were discarded as well because of the steep decay in plasma density and temperature within the arc periphery [16]. The upper point of the stone was located at the height of the cathode surface, so that the sample was facing the inter-electrode gap as the anodic arc discharges evolved. Fig. 1b shows an example of ablated quartz surface after arc plasma treatment. Finally, Fig. 1c displays the characteristic sizes and distances of the experiments, while Fig. 1d shows an image of arc discharge striking next to a quartz sample. The experiments lasted from 5 to 20 s and were recorded using a digital camera at a frame rate of 10-15 frames per second. The view port was equipped with a strongly absorbing optical filter. OES characterization was performed through the view port without filter by means of a StellarNet spectrometer with a spectral range between 191.0 nm and 851.5 nm and a resolution of 0.5 nm at an integration time of 1 ms. Spatially-resolved measurements were performed by aiming at the inter-electrode gap with a black tube of 150 mm in length and small aperture (≈1 mm) attached to the OES probe. The bonding structure of the modified samples was characterized via Raman spectroscopy using a Horiba LabRAM-HR system operated with a 532 nm-laser. The sample images and spectra were taken with a microscopic objective of 50X magnification, yielding a sampling spot size of a few microns.

## 2.2. Fluid model simulations

The transport mechanisms of the heat flux have been studied via fluid simulation approach of the plasma parameters. In the 2-D fluid code employed here, arc discharges between two graphite electrodes 2.8 mm apart, with identical configuration of the arc plasma chamber, are simulated as a simultaneous combination of material evaporation, species reaction and transport processes. Standard Navier-Stokes equations are solved conservatively in order to obtain temperature and densities of species within the chamber. This fluid model has been applied in past studies with variations in approximations and numerical framework [16]. Here, the model is solved in fully compressible flow framework with explicit time integration, in which non-equilibrium chemistry is used [17,18]. The model is implemented in USim platform (Tech-X Corporation). Simulations with initially uniform conditions of 300 Torr He and 300 K have been run with a time resolution of a few nanoseconds, and the fluid parameters were updated every 1 microsecond. Stationary conditions are reached very fast in the proximity of the arc source. In fact, stabilization time is well below 0.1 ms for the analyzed discharge currents. The steady state regime near the arc has been determined by simulating at different time intervals the gas temperature field from a 100 A-arc discharge (not shown here).

## 3. Plasma characterization

### 3.1. Estimation of plasma temperature

Fig. 2a shows the optical emission spectra measured in arc discharges (without quartz samples) run at 60 A (low current) and 150 A (high current). The probed space by OES was restricted to the arc column of the inter-electrode gap, which was between around 2 mm and 3 mm in average length during the experiments. The spectrum of the 150 A-arc is characterized by the emission lines corresponding to the $C_2$ Swan system [13,19,20]. On the other side, the spectrum from the 60 A-arc is less intense overall. It evidences a weaker emission from the molecular $C_2$ lines superposed to a dominant background assigned to continuous thermal emission. The vibrational temperature of the hot gas has been estimated from the Boltzmann relation applied to the intensities of the different $C_2$ transition bands, yielding around 8000 K for the 60 A-arc and roughly 10000 K for the 150 A-arc [13,21]. Note that the error bars in both measurements are around 50%. Such uncertainty is attributed to important gradients of plasma temperature within the sampling volume of the inter-electrode gap. By comparing the total intensities of both spectra, it can be deduced that the arc electromagnetic radiation emission scales with the supplied electric power (see Table 2). Thus, it is plausible to assume that the radiating power absorbed by the quartz crystal is directly related to the flowing current.

The gas temperature measurements, extracted from the OES spectra (Fig. 2a), have been contrasted with the temperature fields simulated for 60 A and 150 A without quartz sample. Fig. 2b shows the corresponding temperature mappings near the arc column. The approximate position of the stone is also indicated. One can observe that the arc column consists of a very hot region, of the order of 5000 K, whereas the arc periphery expands laterally 2 cm in an approximately isothermal region at around 2500 K. The elongated isothermal structure expanding horizontally from the cathode surface is ascribed to the boundary conditions set by this particular electrode configuration. The temperature of that region is well above the quartz melting point (1950 K). The arc core temperature computed here lies within the error bars of the values obtained by OES, although the equivalence between gas temperature and $C_2$ vibrational temperature is a controversial topic [22].

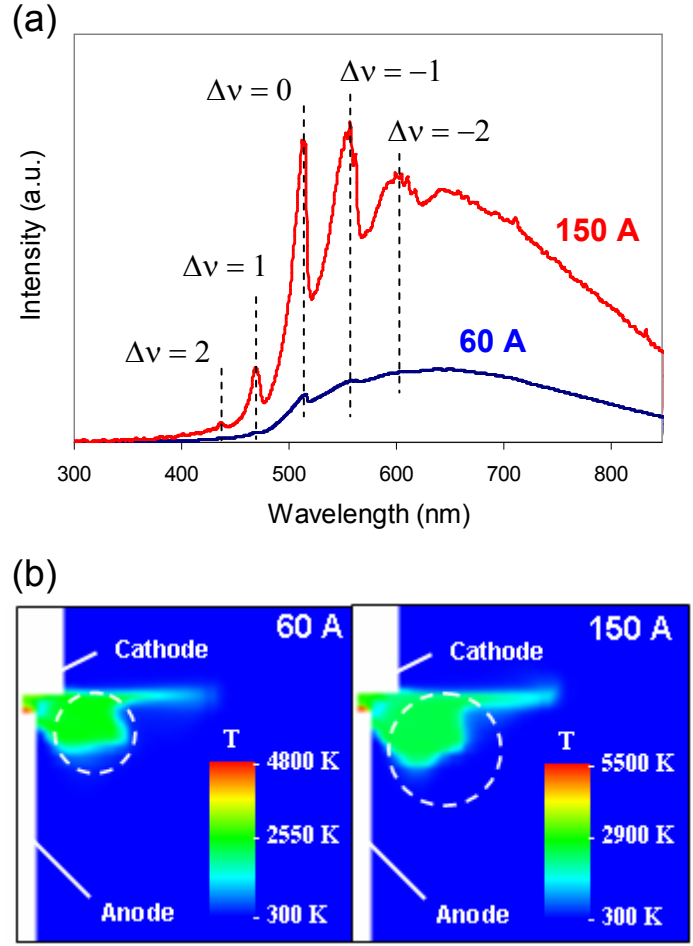

**FIG 2: (a)** Emission spectra measured at the inter-electrode gap in arc discharges held at 60 A and at 150 A. The transitions from the $C_2$ Swan bands, $\Delta v$, are indicated. **(b)** Simulated plots of gas temperature corresponding to arc discharges (without silica spheres) held at 60 A (left) and at 150 A (right), at 0.1 ms after the arc ignition. The circles with dashed lines show the approximate positions of the silica spheres in each experiment. The radii were 5 mm and 7 mm for 60 A and 150 A tests, respectively.

## 3.2. Transport of heat flux

Table 1 displays the gas temperature and gas pressure values calculated for 60, 100 and 150 A in arc current. The considered regions were the centre of the anode surface and the region nominally occupied by the exposed surface of the quartz sphere facing the arc discharge.

**TABLE 1:** Temperatures, $T$, and pressures, $p$, at the onset of the anodic arc (0.1 ms) computed by means of USim simulations. The labels *arc* and *surface* denote the locations where the plasma parameters were calculated.

| Current (A) | $T_{arc}$ (K) | $T_{surface}$ (K) | $p_{arc}$ (Torr) | $p_{surface}$ (Torr) |
|---|---|---|---|---|
| 60 | 4800 | 2500 | 1605 | ≈300 |
| 100 | 5100 | 2550 | 1710 | ≈300 |
| 150 | 5500 | 2550 | 1830 | ≈300 |

The data from Table 1 suggest the presence of important gradients in temperature and pressure, which are translated into respective heat fluxes via radiation, conduction and convection. Although heating contribution by thermal radiation can be dominant at short distances, its quantification is a challenging task because of the requirement of a calibrated OES spectrometer. Besides, the surface emissivity of the irradiated sample is a critical parameter for an accurate prediction of its temperature. The heat flux fraction transported via convection can be qualitatively addressed by simulating the dynamics of carbon species and helium atoms during the gas expansion. Fig. 3 shows the 2-D mappings of the densities of neutral species (C, $C_2$, $C_3$ and He) and ions ($C^+$) for arc simulations at 60 A and 150 A. Both discharges provide structures with similar volumes and shapes. In some cases, the field contours show significant correlation with the temperature distributions plotted in Fig. 2b. Such correlations are discussed next.

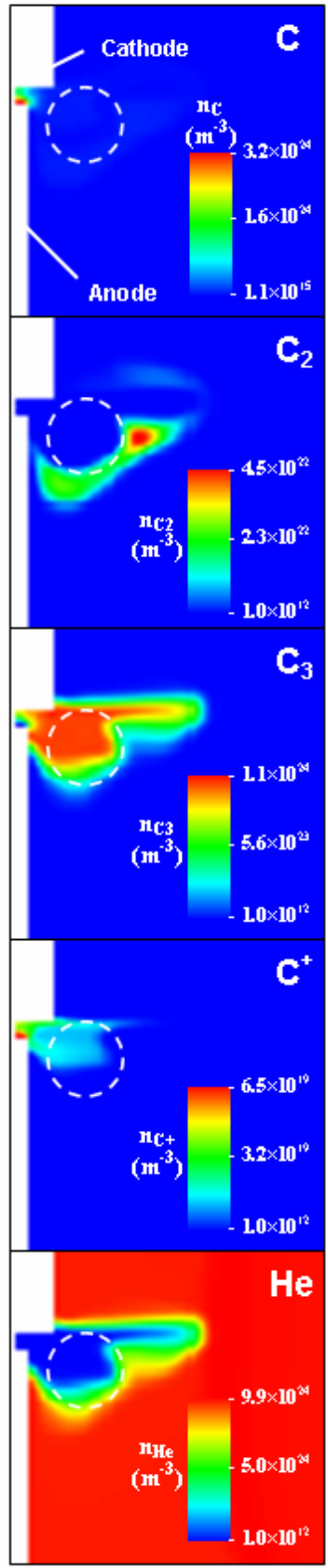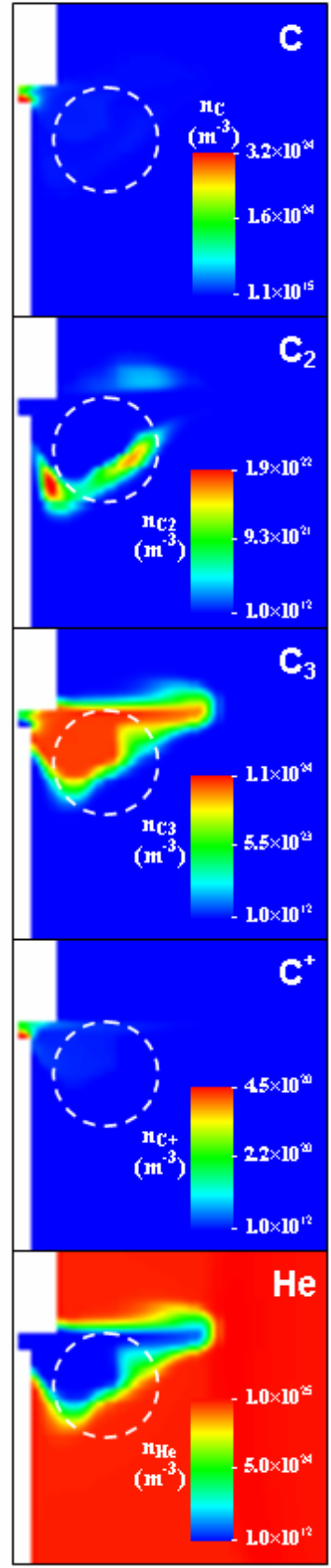

**FIG 3:** 2D mappings of the densities of gas species corresponding to arc discharges (without silica spheres) held at 60 A (left) and at 150 A (right). The data are simulated at 0.1 ms after the arc ignition. The circles with dashed lines show the approximate positions of the silica spheres in each experiment. The radii were 5 mm and 7 mm for 60 A and 150 A tests, respectively.

The 2-D mappings of the $C_3$ fields at the electrode periphery are similar to those of gas temperature, thereby suggesting a correlation between hot gas regions and enhancement of reaction rates. Partially because of its condition of by-product, $C_2$ density is significantly smaller than C and $C_3$ densities. C atoms and $C^+$ ions are dominant at the arc column due to the high plasma temperatures reached in this region, which enhance dissociation and ionization of the carbon species. Finally, the observed depletion of He atoms is consistent with the OES observation of a pure carbon plasma in the arc column (Fig. 2a). Actually, the 2-D profile of the depleted He region is similar to that of gas temperature and of $C_3$. Thus, $C_3$ may be considered as the dominant carbon molecule that displaces helium in the hot gas expansion.

The contribution to the energy flux rate at the crystal surface by conduction, $dq/dt$, can be roughly estimated by means of the Fourier's law assuming isotropic heat flux: $dq/dt = -\kappa \cdot \Delta T/\Delta r$ ,where $\kappa$ is the gas thermal conductivity ($\approx$400 mW/m·K for He) [23], $\Delta T$ is the gas temperature variation ($\approx$2500 K), and $\Delta r$ the radial separation from arc centre ($\approx$3 mm). Here, a spherically symmetric heat propagation is assumed. Top values of the order of 30 W/cm$^2$ are approximated for $dq/dt$, which comprise between 10% and 50% of the total heat flux in the quartz ablation tests discussed later (see Table 2). Such values are higher estimates for the conduction rates because they are calculated using the κ value for He. The real conduction fluxes might be much lower since the C vapour from the anode displaces the He atoms (Fig. 3).

In summary, the simulation results suggest that convection is an important heating mechanism of the surrounding gas. However, the gas temperature undergoes a small increase of 2% as the arc current is increased from 60 A (2500 K) to 150 A (2550 K) (Table 1), thereby indicating that current variations do not impact the convective heating to a great extent in such current interval. On the contrary, the thermal radiation flux shows a stronger dependence with the supplied power, as illustrated by the different intensities of optical spectra depicted in Fig. 2a. Thermal radiation is, therefore, the transport mode of heat flux better connected to the discharge current. Actually, electromagnetic radiation constitutes a relevant heating mechanism in solid surfaces.

## 4. Ablation experiments

### 4.1. Model of arc plasma-crystal interaction

From a macroscopic point of view, the energy source of the arc discharge is based on heat dissipation of the current by Joule effect. The energy generated by the arc plasma expands through the helium atmosphere by radiation, conduction and convection mechanisms. Then, the total energy flux collides with the quartz crystal by increasing its temperature and promoting melting and vaporisation of the solid material. Moreover, the influx of carbon species generated by ablation of the graphite anode overlaps with the absorbed heating power, and any eventual deposition could modify the surface properties of the quartz sample.

In view of the characteristics of the arc discharge, which have been approached by optical measurements and numerical simulations, the analysis of the ablation experiments has been performed by considering the following *hypotheses*:

- Isotropic expansion of heat flux: The heat flux is transported basically by radiation and convection with an energy flow emerging punctually from the hot arc source. Heat transport by conduction is marginal. As observed from the temperature mappings calculated by the arc discharge simulations, as well as obtained from light emission profiles reported elsewhere [13,21], such energy flux evolves roughly with an isotropic expansion pattern near the arc core. Therefore, the geometrical extinction factor $1/r^2$ applies to the heat flux observed at a small distance $r$ from the arc core.

- Exponential decay of heat flux: The decay of the heat flux with the distance from the emitting arc, $r$, due to gas collision processes is modulated by the exponential factor $\exp(-r/\lambda)$. Such a behaviour is based on the profiles of carbon growth rate obtained earlier by means of fast probe experiments [24]. In past studies, the exponential term $\lambda$ was estimated to span between 1.5 mm and 3.5 mm. Here, the middle value 2.5 mm has been chosen for the sake of simplicity. All loss mechanisms of the energy flux in gas interaction are enclosed in such exponential factor.

- Negligible deposition onto hot quartz surface: The sticking probability of the incident carbon atoms onto the ablating quartz is very small. The high temperature on the sample surface near the arc (gas temperature ≈2500 K) hinders the solidification of carbon. Indeed, coagulation of carbon species submitted to such a thermal load is not possible, as demonstrated by arc sampling experiments with a fast tungsten probe and similar arc parameters [25]. Therefore, the deposition of carbon onto the ablating surface of the stones is assumed to be negligible.

Any residual carbon detected post-treatment on the arc-facing surface is attributed to graphitic or amorphous carbon deposition during the short time interval of arc extinction.

In the following section, the arc plasma ablation experiments on quartz stones are discussed. The quartz samples are placed near the inter-electrode space, leaving their surface a gap of 2 mm with the anode surface. The aforementioned assumptions are considered plausible in order to correlate the evaporation rate of the solid with the absorbed power of heat flux.

**4.2. Determination of specific energy**

Table 2 lists the tested samples together with the arc parameters and the resulting ablation rates. The reference, pristine quartz sample is termed Sample 0. Such rates, which increase monotonically with the arc current, are obtained from the mass difference of each sample measured before and after arc treatment using a microbalance. Before measuring the weight of the treated stones, any residual carbon coating was carefully removed with a cotton swab wetted with ethanol. The power density absorbed by the stone samples, $W_{abs}/A$, is calculated from the geometric parameters defined in the sketch from Fig. 1c. Basically, the absorbed power density is the estimated power projected to the exposed stone surface with area $A$. As discussed in the previous sections, $W_{abs}$ was estimated by considering isotropic power emission from the arc core, which was centred on the anode top. Besides geometric extinction of the heat flux, radial collisional losses associated to gas interaction have been also included. As a result, the incident flux $W_{abs}/A$ is affected by an exponential decay $\exp(-r/\lambda)$ with $r \approx \lambda$. The so calculated power densities absorbed by the quartz crystals range between 30 and 250 W/cm$^2$.

**TABLE 2:** List of treated quartz samples together with arc discharge parameters and results. All stones were held at 2 mm distance from the anode lateral surface. The absorbed power density was estimated by considering gas collision losses and isotropic power emission emerging from the arc core.

| Sample nr. | Radius (mm) | Current (A) | Voltage (V) | Power (W) | Time (s) | Ablation rate (mg/s) | Absorb. power dens. (W/cm$^2$) | Penetration rate (cm/h) |
|---|---|---|---|---|---|---|---|---|
| 1 | 5 | 30 | 25 | 750 | 19 | <0.1 | 31 | <0.2 |
| 2$^a$ | 5 | 220 (max) | 50 (max) | 900 | 16 | 0.9 | 37 | 2.1 |
| 3 | 5 | 60 | 25 | 1500 | 7 | 2.2 | 63 | 5.0 |
| 4 | 5 | 75 | 32 | 2400 | 7 | 3.8 | 100 | 8.7 |
| 5 | 5 | 100 | 38 | 3800 | 7 | 6.5 | 160 | 15.0 |
| 6 | 7 | 125 | 44 | 5500 | 6 | 7.1 | 160 | 9.0 |
| 7 | 7 | 150 | 55 | 8300 | 5 | 6.7 | 240 | 8.5 |

$^a$ Pulsed arc discharge conducted at 2 Hz with 10% of duty cycle (pulse duration: 50 ms)

The penetration rate, $P$, is defined as the ratio between $W_{abs}/A$ and S.E. of the treated material [9]:

$$P = \frac{1}{S.E.} \cdot \frac{W_{abs}}{A}. \qquad (1)$$

From Eq. (1), it is straightforward to prove that the penetration rate is proportional to the ablation rate of the quartz sample, $R$, as $P=R/(A\rho)$. Such expression is used to calculate $P$ from the measured values of $R$, all listed in Table 2, which show a maximum of 15 cm/h at a set current of 100 A (3800 W). The measured mass density of quartz is $\rho=2.65$ g/cm$^3$. The exposed areas, $A$, are 59 mm$^2$ and 105 mm$^2$ for the quartz crystals with respective 10 mm and 14 mm in diameter. The exposure of the quartz spheres to the arc discharge induced deformation of the samples, together with a phase change from crystalline to glassy silica, as demonstrated by Raman spectroscopy (not shown here).

Fig. 4 shows the evolution of the arc energy penetration rate as a function of the power density absorbed by the stone. As indicated in Eq. (1), the inverse of the slope from the linear fit provides the specific energy for quartz. The value of S.E. obtained in the present tests is approximately 40 kJ/cm$^3$, which is comparable to the typical values measured in mineral rocks (26 kJ/cm$^3$ for granite and basalt, 32 kJ/cm$^3$ for limestone) [9]. The data points obtained at 30 A (Sample 1), 125 A (Sample 6) and 150 A (Sample 7) in arc current are excluded from the linear fit analysis because they correspond to different arc regimes. In particular, the 30 A-discharge is developed at the transition from a constricted arc regime, in which the ablation rate is negligible. On the other hand, the discharges held at high currents promote very fast ablation of the graphite anode [13]. This scenario may contribute to a non-negligible carbon flux onto the sample, which competes with the ablation flux of the stone itself. Such a carbon layer interferes also with the plasma energy influx to the sample, thereby invalidating the assumption of weak carbon deposition stated earlier. Also, intense arcs yielded to uncontrolled migration of the plasma region, which ended up in an irregular, non-reproducible ablation of the sample.

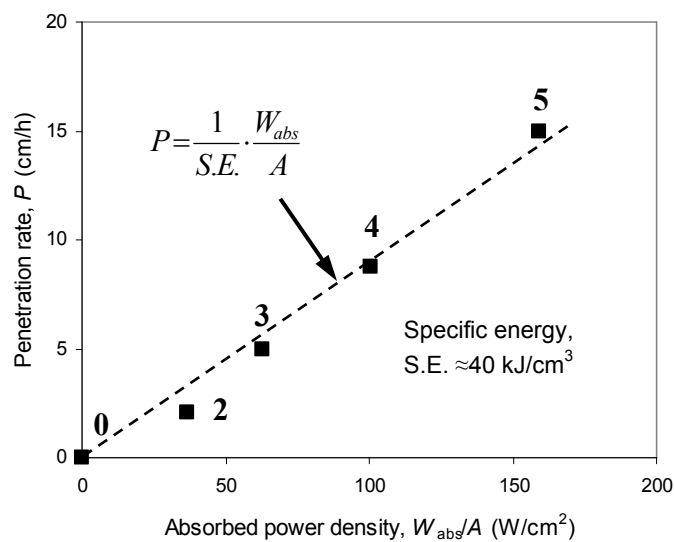

**FIG 4:** Penetration rate of arc plasma energy on quartz at different absorbed power densities. The sample nr. is placed near each point. The specific energy, S.E., is derived from the linear fit of Eq. (1) to the experimental data.

## 5. Discussion

This study shows that the penetration rate of plasma energy into quartz scales linearly with the absorbed power density. Helium was used as background gas because it is a standard in atmospheric arc process research. Moreover, the technology can be optimized for its transfer to realistic environments, for example water vapour atmosphere as used in high-power plasma torches [11]. A rate of up to 15 cm/h was achieved within the investigated power interval. The arc plasma drilling method can be competitive with other techniques, like mechanical drilling and MMW deep drilling, upon achievement of penetration rates overcoming several m/h [9]. For this, power density values of several kW/cm$^2$ are required. Such power levels are feasible by upgrading the hypothetical arc discharge instrument into a more robust system with increased heat capacity (wider graphite anodes). Hence, the integrity of the arc electrodes could be preserved for a longer time thanks to minimization of the anode erosion rate. The modulation of arc current waveform in periodical pulses has also proven helpful in extending the lifetime of the electrodes [13].

An important issue in arc plasma technique towards rock drilling performance is the omni-directional energy emission from the discharge. Such a difficulty can be addressed by enhancing both focus and arc stability via magnetic fields and/or additional plasma torches. Indeed, an arc discharge drilling machine could be scaled up into a large plasma source which includes a cluster of single arc sources. Interestingly, there is in principle no limitation for the working power set to the instrument as long as the electrode system is not degraded. Such liberty is not possible in the case of directed MMW drilling, because gas breakdown is predicted to occur generally at power densities higher than 50 kW/cm$^2$, which restricts the penetration rate to about 70 m/h in that particular technique [9]. The fact that anodic arc discharge source is already in plasma state constitutes a distinct advantage of arc plasma in front of the other established techniques. Focusing the arc power is also a necessary step. Actually, one can estimate the rate of penetration provided an optimal arc driller with the plasma power fully oriented in the forward direction. As a reference, focused arc operation at a supplied power of 100 kW on quartz (S.E.≈40 kJ/cm$^3$) would yield a penetration rate of approximately 90 m/h when drilling an area of 1 cm$^2$ (Eq. (1)). A larger drilling area can be achieved by increasing the size of the plasma source and by clustering a number of plasma sources. This benchmark justifies the competitive performance of arc plasma ablation compared to the traditional drilling techniques.

## 6. Conclusions

Ablation processes of quartz crystals have been systematically approached with atmospheric arc discharges using graphite electrodes for the first time. The rate of penetration of the absorbed power is proportional to the ablation rate, and it shows a linear trend with the power density yielding an effective specific energy of approximately 40 kJ/cm$^3$. Optical characterization and fluid simulations of the arc discharge revealed that quartz heating from the high temperature plasma is governed by radiation and convection fluxes. A maximal value of penetration rate of 15 cm/h is obtained for an incident power density of around 150 W/cm$^2$, thereby demonstrating that arc plasma technology is promising for mineral cutting and drilling applications. Extension of electrode lifetime and optimization of power guiding constitute the main challenges to address.

## Acknowledgments

No funding was received for conducting this study. The authors gratefully acknowledge the support by Dr. M.N. Kundrapu (Tech-X Corp.) for providing the USim software platform.

## References

[1]     Iijima S 1991 Helical microtubules of graphitic carbon *Nature* **354** 56

[2]     Boxman R L, Sanders D M, Martin P J 1995 *Handbook of Vacuum Arc Science and Technology* (Park Ridge: Noyes Publications)

[3]     Anders A 2008 *Cathodic Arcs: From Fractal Spots to Energetic Condensation* (Springer-Verlag, New York: Springer-Verlag)

[4]     Shashurin A and Keidar M 2015 Synthesis of 2D materials in arc plasmas *J. Phys. D: Appl. Phys.* **48** 314007

[5]     Keidar M and Beilis I I 2018 *Plasma Engineering* 2nd edn (London: Elsevier)

[6]     Zhang J, Sugioka K and Midorikawa K 1998 High-speed machining of glass materials by laser-induced plasma-assisted ablation using a 532-nm laser *Appl. Phys. A* **67** 499-501

[7]     Qin S -J and Li W J 2002 Micromachining of complex channel systems in 3D quartz substrates using Q-switched Nd:YAG laser *Appl. Phys. A* **74** 773–777

[8]     Kazi A, Riyaz M, Tang X, Staack D and Tai B 2020 Specific cutting energy reduction of granite using plasma treatment: A feasibility study for future geothermal drilling *Procedia Manufacturing* **48** 514-519

[9]     Woskov P and Cohn D 2009 *Annual Report 2009: Millimeter Wave Deep Drilling For Geothermal Energy, Natural Gas and Oil MITEI Seed Fund Program* (Cambridge: MIT Plasma Science & Fusion Center) http://hdl.handle.net/1721.1/93312. Accessed 12 November 2020

[10]    Dzur B 2007 Plasma processing of concrete and related materials *High Temp. Mater Process.* **11** 493-503

[11]    Stefanov P, Galanov D, Vissokov G, Paneva D, Kunev B and Mitov I 2008 Study of Nanodispersed Iron Oxides Produced in Steel Drilling by Contracted Electric-Arc Air Plasma Torch *Plasma Sci. Technol.* **10** 352

[12]    Bazargan M, Gudmundsson A and Meredith P I 2017 Feasibility of Using Plasma Assisted Drilling in Geothermal Wells *Proc. of the 4th Sustainable Earth Sciences Conference, Sep. 2017,* 1-5

[13]    Corbella C, Portal S, Zolotukhin D B, Martinez L, Lin L, Kundrapu M N and Keidar M 2019 Pulsed anodic arc discharge for the synthesis of carbon nanomaterials *Plasma Sources Sci. Technol.* **28** 045016

[14]    Corbella C, Portal S, Saadi M A S R, Solares S D, Kundrapu M N and Keidar M 2019 Few-layer flakes of Molybdenum Disulphide produced by anodic arc discharge in pulsed mode *Plasma Res. Express* **1** 045009

[15]    Fang X, Shashurin A and Keidar M 2015 Role of substrate temperature at graphene synthesis in an arc discharge *J. Appl. Phys.* **118** 103304

[16]    Kundrapu M and Keidar M 2012 Numerical simulation of carbon arc discharge for nanoparticle synthesis *Phys. Plasmas* **19** 073510

[17]    Kundrapu M, Averkin S, Stoltz P and Keidar M 2018 Software for plasma device simulations: Arc plasma sources *Plasma dynamics and Lasers Conference 2018* (p. 2940)

[18]    M. Kundrapu, A. Chap, M. De Messieres, C. Corbella and M. Keidar 2019 Two-Temperature Simulation of Subatmospheric Arc Discharge *IEEE Pulsed Power and Plasma Science, PPPS 2019, Orlando, FL*.

[19] Danylewych L L and Nicholls R W 1974 Intensity measurements on the $C_2$ ($d^3\Pi_g - a^3\Pi_u$) Swan band system *Proc E Soc Lond A* **339** 197-212

[20] Li J, Kundrapu M, Shashurin A and Keidar M 2012 Emission spectra analysis of arc plasma for synthesis of carbon nanostructures in various magnetic conditions *J Appl Phys* **112** 024329

[21] Corbella C, Portal S, Kundrapu M N and Keidar M 2020 Energy considerations regarding pulsed arc production of nanomaterials *J. Appl. Phys.* **128** 033303

[22] Carbone E, D'Isa F, Hecimovic A and Fantz U 2020 Analysis of the $C_2$ ($d^3\Pi_g - a^3\Pi_u$) Swan bands as a thermometric probe in $CO_2$ microwave plasmas *Plasma Sources. Sci. Technol.* **29** 055003

[23] *Thermophysical Properties of Materials For Nuclear Engineering: A Tutorial and Collection of Data*. IAEA-THPH, IAEA, Vienna, 2008. ISBN 978–92–0–106508–7

[24] Corbella C, Portal S, Rao J, Kundrapu M N and Keidar M 2020 Tracking nanoparticle growth in pulsed carbon arc discharge *J. Appl. Phys.* **127** 243301

[25] Fang X, Shashurin A, Teel G and Keidar M 2016 Determining synthesis region of the single wall carbon nanotubes in arc plasma volume *Carbon* **107** 273–280